# On the Interaction between Software Engineers and Data Scientists when building Machine Learning-Enabled Systems

Gabriel Busquim, Hugo Villamizar, Maria Julia Lima, and
Marcos Kalinowski

Pontifical Catholic University of Rio de Janeiro, Brazil
{gbusquim,hvillamizar,kalinowski}@inf.puc-rio.br,
mjulia@tecgraf.puc-rio.br

**Abstract.** In recent years, Machine Learning (ML) components have been increasingly integrated into the core systems of organizations. Engineering such systems presents various challenges from both a theoretical and practical perspective. One of the key challenges is the effective interaction between actors with different backgrounds who need to work closely together, such as software engineers and data scientists. This paper presents an exploratory case study that aims to understand the current interaction and collaboration dynamics between these two roles in ML projects. We conducted semi-structured interviews with four practitioners with experience in software engineering and data science of a large ML-enabled system project and analyzed the data using reflexive thematic analysis. Our findings reveal several challenges that can hinder collaboration between software engineers and data scientists, including differences in technical expertise, unclear definitions of each role's duties, and the lack of documents that support the specification of the ML-enabled system. We also indicate potential solutions to address these challenges, such as fostering a collaborative culture, encouraging team communication, and producing concise system documentation. This study contributes to understanding the complex dynamics between software engineers and data scientists in ML projects and provides in-sights for improving collaboration and communication in this context. We encourage future studies investigating this interaction in other projects.

## 1 Introduction

Integrating Machine Learning (ML) components into existing systems has increased as companies seek to leverage vast amounts of data to enhance the business outcomes of their software products. In this paper, we refer to these systems as ML-enabled systems. Typically, the ML component is only a small part of a larger system [1], which usually comprises other components for data collection, model consumption, and infrastructure requirements.

This transition from developing traditional software systems to those integrated with ML components introduces new challenges from the viewpoint of Software Engineering (SE). The development of ML-enabled systems often involves completely separate workflows, as well as different actors [2]: data scientists build ML models while engineers must deploy and integrate them with other services. An ineffective interaction between team members can cause ML mismatches capable of harming the system [3]. This scenario raises the question of whether proper alignment and communication between the actors occurs and how they share responsibilities when developing ML-enabled systems.

Following the guidelines by Runeson *et al.* [4] for case study research in software engineering, we tackle this issue by conducting an exploratory case study focused on two key roles within ML projects: software engineers and data scientists. The selected case concerns an ML-enabled system for Online Dispute Resolution (ODR) created to help parties settle legal disputes in the state of Rio de Janeiro. Beyond describing the team and system context, we conducted semi-structured interviews with four experienced team members, two software engineers and two data scientists, to understand their current interactions, collaboration dynamics, and problems in ML projects. To this end, we asked practitioners about activities covering the development process end-to-end. Our questions range from defining requirements to analyzing data and integrating the ML model with the rest of the system. We transcribed and analyzed the interviews using reflexive thematic analysis [5,6], one of the Thematic Analysis (TA) family methods. This research approach guided us while analyzing the data and finding patterns among the interviewees' points of view.

We divided our findings into five main categories: **requirements**, **planning**, **data management**, **model management**, and **team interaction**. We illustrate the participants' perceptions and the main improvement opportunities they noted for each category. Respondents expressed several challenges regarding their tasks and current collaboration practices. For example, data scientists and software engineers were not always aware of each other's activities, which led to inaccurate planning and errors when integrating the model with the rest of the system. Even though they viewed their relationship positively, they recognized that a more efficient collaboration could have prevented the late discovery of errors in the system. Our main contribution with this work is highlighting the importance of having well-defined responsibilities and collaboration procedures inside teams developing ML-enabled systems. By reporting challenges faced by professionals, we seek to instigate practitioners to evaluate their collaboration practices since the beginning of the project.

## 2   Background and Related Work

### 2.1   Challenges in Building ML-enabled Systems

Villamizar *et al.* [7] define ML-enabled systems as software systems with an ML component. The development of ML-enabled systems presents several challenges that can significantly impact the interaction between team members. This is the case especially for software engineers and data scientists, who often share responsibilities for handling data and deploying models [8]. For example, designing an appropriate architecture for these systems is not trivial, as the team must evaluate factors such as model performance degradation, uncertainty management, and proper integration between the model and other system components [9].

Furthermore, requirements engineering practices for non-ML software development are not entirely applicable when developing systems with an ML component [10]. There is a typical lack of requirements specifications for such systems [11] that provide a clear definition of the input data, expected model outputs, and how the ML component should integrate into the larger system [7]. Without these specifications, data scientists may create models with assumptions that software engineers are unaware of, leading to integration issues when transitioning from development to production.

The different backgrounds of data scientists and software engineers can also impact their interactions. While data scientists may have strong mathematical and statistical skills [12], software engineers have expertise in programming, software design, and system architecture. This diversity can lead to variations in problem-solving approaches. In addition, their cultural differences can also play an important role. While the tasks performed by data scientists revolve around experimentation and dealing with the uncertainty of unpredictable results [2], software engineers often adhere to structured development methodologies. These cultural disparities can cause barriers in a collaborative environment.

### 2.2  Communication and Collaboration in ML-enabled Systems

Amershi *et al.* [13] presented a case study with Microsoft software teams to gather best practices for ML engineering. Results showed how respondents consistently cited collaboration as a challenge. Communication and collaboration are also mentioned in papers examining the role of data scientists. Kim *et al.* [12] presented a survey with data science employees at Microsoft to uncover the challenges they face. Some were related to team communication, such as effectively transmitting insights to leaders and achieving agreement among all stakeholders.

Specifically focusing on collaboration, Zhang *et al.* [14] conducted a survey on how data science workers, including data scientists and software engineers, collaborate. The results depicted how data scientists were engaged throughout all steps of data science projects, while software engineers were more involved in core technical activities, such as acquiring data for the model. Lewis *et al.* [3] studied the consequences of ML mismatches between data scientists, software en- gineers, and operations staff developing ML-enabled systems. They interviewed practitioners to understand examples and recommendations for avoiding these problems. Results showed that most mismatches were related to incorrect as- sumptions about the model. They also refer to a lack of model specifications and test cases for integration testing. These issues are directly related to the interaction between data scientists and software engineers.

More recently, Mailach and Siegmund [15] investigated sociotechnical challenges for bringing ML-enabled software into production. They identified chal-lenges related to organizational silos, especially between the data science and software engineering teams. The paper reported tension and communication is-sues when the teams collaborated, which led to production delays. Nahar *etal.* [16] focused on identifying challenges and recommendations for the inter- action between software engineers and data scientists. They mapped several collaboration points between the two actors, from project planning to product-model integration. As in Mailach and Siegmund's study [15], participants also reported problems with data scientists working in isolation and communication issues between them and software engineers.

When discussing the state of the art, Nahar *et al.* [16] mentioned they were unaware of other studies examining challenges between software engineers and data scientists. With our work, we intend to expand the literature on this topic and provide additional insights through a case study strategy. Hence, differently from Nahar *et al.*, who covered perspectives from multiple teams from different organizations, we qualitatively analyzed a selected case, providing its context and conducting thematic analysis. Beyond examining the collaboration between data scientists and software engineers, the case study strategy also allowed us to qualitatively understand the responsibilities these actors had during the execution of the selected case project.

## 3   Case Study Design

We conducted a case study to enhance our comprehension of the interaction and collaboration dynamics between software engineers and data scientists. Here- after, we describe its design following the guidelines by Runeson *et al.* [4].

### 3.1   Goal and Research Questions

The goal of this study, described following the Goal-Question-Metric (GQM) template for goal definition [17], can be seen in Table 1. From this goal, we derived the following research questions.

**Table 1.** Case Study Goal

| | |
|---|---|
| **Analyze** | the interaction between software engineers and data scientists |
| **for the purpose of** | characterization |
| **with respect to** | responsibility sharing and collaboration |
| **from the point of view of** | experienced software engineers and data scientists |
| **in the context of** | a large ML-enabled system project for Online Dispute Resolution (ODR) to help settle legal disputes. |

*RQ1: How do software engineers and data scientists share responsibilities when developing an ML-enabled system?*

This research question focuses on how responsibilities are shared, providing insights into the task allocations and synergies that contribute to the successful creation of ML-based solutions. To answer *RQ1*, we evaluated the participation of software engineers and data scientists in multiple stages of the ML-enabled system's creation, such as during the system's design and model development. For each activity, we mapped the actors and if any collaboration happened.

*RQ2: How do software engineers and data scientists collaborate when devel- oping an ML-enabled system?*

This question focuses on the collaboration between software engineers and data scientists during the development of ML-enabled systems. It seeks to un- cover the nature of their interactions, communication methods, and joint efforts, contributing to understanding the collaborative processes. To this end, we asked participants about their perceptions of how this interaction unfolded inside the team. We encouraged them to highlight challenges and improvement possibil- ities, which we used to formulate recommendations for other teams building ML- enabled systems.

## 3.2   Case and Subject Selection

The selected case concerns an Online Dispute Resolution (ODR) system project. It was created to help parties settle legal disputes in Rio de Janeiro. The system uses ML to generate settlement agreements for cases with low legal complexity, therefore avoiding litigation. We chose to focus on this particular project because it is centered around the development of an ML-enabled system, aligning with the scope of our intended investigation. Furthermore, we had easy access to project participants and the complete system documentation.

The project started in 2021 inside PUC-Rio's Tecgraf Institute through a partnership with the Rio de Janeiro State Court. After applying the Lean In- ception methodology [18], the team defined the product's main functionalities. Given the system's goal, developing an ML component to aid in dispute res- olution was considered an interesting choice. This led to the incorporation of data scientists into the team, which also began participating in meetings to un- derstand business rules and discuss model characteristics. For the system's first version, the team partnered with an electric power company and established their focus on disputes involving consumer complaints directed to this company. The company representatives then developed external APIs that the system would consume to obtain all the data required by the model.

**Process and Team Configuration.** The project follows the Scrum framework with sprints of two weeks. Ceremonies include sprint planning, daily meetings, sprint review, and sprint retrospectives. The team responsible for developing the system is multidisciplinary. It comprises a project manager, domain experts, UX designers, data scientists, and software engineers. All team members par- ticipated in meetings to understand business rules and discuss solution ideas. Customer representatives also attended these meetings to ensure decisions fol- lowed their expectations. Besides providing requirements, they also evaluated the team's deliveries through release versions made available by the software engineers every two months. With respect to the target roles, the team comprises six software engineers and two data scientists, considered part of two separate squads. Each squad has its tasks, as well as its own planning and daily meetings. However, the teams share the same product owner.

**Architecture and ML Component.** Figure 1 provides an overview of the system's architecture. Users have access to the system's functionalities through a web application that communicates with back-end services through a REST API. The back-end architecture is based on microservices, with each service having a single responsibility. The services communicate both synchronously and asynchronously. Synchronous communication happens through REST APIs, while asynchronous communication occurs via message queues.

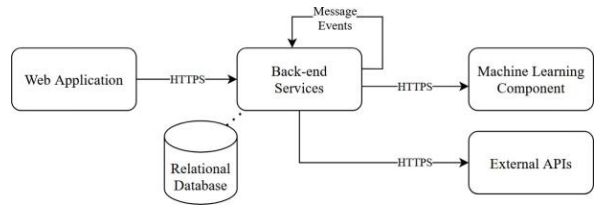

**Fig. 1.** Simplified System Architecture

One of the system's back-end services communicates with the ML component through a REST API. The model's input consists of data entered by users on the web system and complementary data obtained from the external APIs. As output, the model returns whether it can generate a settlement agreement. If the result is positive, the model returns all agreement parameters. If it is negative, the model returns why it could not create an agreement.

The model consists of a decision tree with a set of fixed rules, defined by customer representatives and the partner company, that must be validated before the system can generate a settlement agreement. These rules were created to restrain the model's possible outputs and improve transparency. Having verified all rules with a positive outcome, the model evaluates data from other previously resolved disputes. After selecting and analyzing the most similar disputes, the model defines the ideal value for each settlement agreement parameter, such as the value for compensating moral damages. The text classification method behind the model's functionality is described in the work of Coelho *et al.* [19].

### 3.3   Data Collection

We formulated our interview questions based on the work of Villamizar *et al.* [7], which offers a conceptual diagram that models tasks and related concerns typically faced by different stakeholders in ML projects. Using such diagram, we initially mapped tasks associated with the infrastructure perspective involving either a software engineer or a data scientist, presented in Table 2.

**Table 2.** Tasks of the Infrastructure Perspective

| Task | Description |
|------|-------------|
| Update the Model | Involves specifying how the ML-enabled system can continuously learn from new data. |
| Make the Model Available | Concerns defining how the model will be consumed, e.g., through a web endpoint. |
| Observe the Model | Concerns determining how model performance and results will be monitored. |
| Store the Model | Involves defining where the ML artifacts, such as models and scripts, will be stored. |
| Integrate the Model | Addresses how communication between components is established to provide functionality for the ML-enabled system. |

We also investigated tasks from perspectives outside the system's infrastructure, described in Table 3. We did this to have a broader view of the responsibilities of software engineers and data scientists inside the project. Given that all perspectives may instigate interaction between a data scientist and a soft- ware engineer, we designed our interview script to investigate how participants handled these tasks in the context of the project. Specifically, we had questions about (i) the interviewee's participation in each task, (ii) the interaction with a data scientist or software engineer on that task, (iii) the perceived difficulties or improvement opportunities during task execution, and (iv) the documentation originated by that task. We used this interview design to guide the discussions we had with the participants while allowing them to share their thoughts and insights freely. We recorded all interviews and, to transcribe them, we used Google Cloud's Speech-to-Text API[1].

**Table 3.** Evaluated Perspectives

| Perspective | Description |
|-------------|-------------|
| System Objectives | Involves understanding the problem to be solved by the ML-enabled system and defining the model's goals. |
| User Experience | Involves designing an appropriate interaction between the user and the model. |
| Data | Addresses how data is obtained and analyzed to build the model. |
| Model | Concerns defining the model's inputs and outputs and evaluating its performance. |

### 3.4   Analysis  Procedure

After acquiring all text files, we analyzed each transcription and made corrections while listening to the recordings. We also removed direct references to employee names to guarantee anonymity. The revised interview transcriptions can be found in our online open science repository[2].

---

[1] https://cloud.google.com/speech-to-text

For analyzing the data, we followed the guidelines for reflexive thematic analysis (RTA) defined by Braun and Clarke [5, 6]. Although RTA is widely used in psychology research, studies have shown that it can be applied in other fields, such as software engineering [20] and human-computer interaction [21]. We de- cided to use RTA in our research since it allows us to engage analytically with the data. In other types of TA methodologies, such as coding reliability approaches, the analysis provides summaries of what was said about a particular topic [22]. In our case, we were interested in finding and interpreting patterns inside the data to fully understand the scenario illustrated by our participants and extract the main challenges they reported. Following the recommendations of Brown and Clarke [22], we did not consider using grounded theory due to the small size of our sample and the fact we do not have the goal of developing a theory.

The first phase of RTA is to familiarize with the data, which we did while re- viewing the transcriptions and listening to the recordings. After that, we started the coding process. With this process, we aim to group together different data components so that all information covering a given topic is in the same category. To do this, we first read each transcript thoroughly. Then, for each relevant text fragment, we create a code. As we keep reading, we either assign more sentences to one of the codes or create a new one. We followed an inductive approach for coding, where codes are developed using the data itself as a starting point.

With the codes defined, we grouped them into themes. To find them, we looked for similarities between the codes. Themes should be objective and un- derpinned by a central concept. They must contain useful information about the dataset, directly addressing at least one research question. Following RTA recommendations, we iteratively refined the themes until they met these criteria.

## 4   Case Study Results

### 4.1   Participant Characterization

The participants verbally agreed to participate voluntarily in the study and have their interviews recorded. All subjects identified as male and hold a master's or a doctorate degree. Table 4 shows the roles, education level, and years of work experience for each one.

### 4.2   Results

We summarized our findings into five main categories: **requirements**, **planning**, **data management**, **model management**, and **team interaction**. We included direct quotes and paraphrased statements from the practitioners to support the analysis and interpretations.

---

[2]  https://doi.org/10.5281/zenodo.10035304

**Table 4.** Demographic Data about the Respondents

| Participant ID | Role | Education Level | Years of Experience |
|---|---|---|---|
| DS1 | Data Scientist | Doctorate | 8 |
| DS2 | Data Scientist | Doctorate | 8 |
| SE1 | Software Engineer | Master's degree | 11 |
| SE2 | Software Engineer | Master's degree | 12 |

**Requirements.** An overview of the case study findings related to requirements can be seen in Figure 2. An explanation for each one follows.

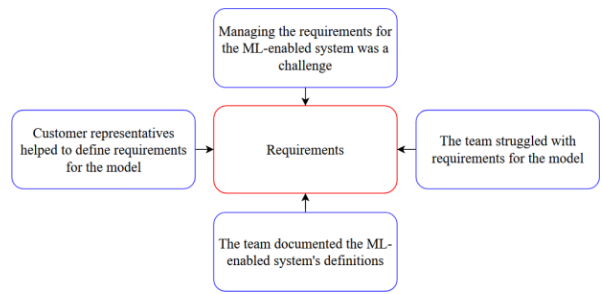

**Fig. 2.** Findings for the Requirements category

**Managing the requirements for the ML-enabled system was a chal- lenge.** Participants emphasized that requirements constantly changed. DS1 pro- vided an example: "*In the beginning, we had defined that the model would be as flexible as possible. We realized during later meetings this would not be well ac- cepted, as it would make the model's results less predictable.*"

**Customer representatives helped to define requirements for the model**. SE1 gave examples of their participation: "*I noticed that customer rep- resentatives could actively suggest model parameter adjustments. Another topic they discussed was keeping information about the model's operation private from end users. This was done to prevent them from learning how to manipulate the model in their favor.*" DS2 also recognized the importance of customer involve- ment, mentioning that he felt like customer representatives could have partic- ipated more: "*We had difficulties because we did not include more customer representatives when we defined the product's concepts. They could have helped us by making decisions. Instead, we made decisions internally. We had to revisit some of these decisions later, while we were lucky not to in others.*"

**The team struggled with requirements for the model**. Data scientists mentioned that model requirements were unrealistic and unclear at the beginning of the project. DS2 stated: "*The requirements were abstract, like 'the model needs to be fast' or 'the system needs to be easy to use.' There was a misalignment between what was desired and what was possible, which led to many meetings.*"

**The team documented the ML-enabled system's definitions** and recognized the importance of doing so. DS2 explained: "*Each model definition was documented through presentations we did in meetings to showcase what our team was proposing. The architecture of the model was also described in a document.*" DS1 highlighted the importance of documenting each meeting: "*We created a flowchart with all the rules the model considered and documented the meetings through minutes. We even had an episode where it was necessary to resort to these minutes to prove that the team had made certain decisions in a previous encounter.*" DS1 also mentioned how these documents helped him learn about the project when joining the team: "*Reports were developed at the beginning of the project [...]. These documents helped me understand the business faster.*"

**Planning.** An overview of the findings related to planning is provided in Figure 3. An explanation of the results that emerged from the analysis follows.

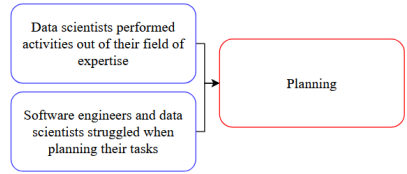

**Fig. 3.** Findings for the Planning category

**Data scientists performed activities out of their field of expertise**, such as eliciting requirements for the system. DS2 explained: "*Our team was responsible for understanding the entire business flow and legal procedures so that we could build the model. Someone else could have done this survey and delivered the requirements to us.*" The data science team also developed the model consumption API. In DS2's view, this should have been done by the software engineers: "*We were a research team, not a development team. Still, we needed to develop versions and generate specifications for the model. Our team was responsible for developing and maintaining the model consumption API. This responsibility could have been given to the software engineering team.*"

**Software engineers and data scientists struggled when planning their tasks**; they tried to plan their activities separately, only communicating when necessary. SE2 explained this process: "*We created a REST API to allow the model integration with the system. We defined a communication interface for the API, and then each team did its part. It was outside the data science team's interest to understand how we stored the data as long as this service existed.*". Nevertheless, some participants were unhappy with this decision, especially with the coordination between the two teams. Each team had its own goals for each sprint, and dependencies between them were not always correctly mapped. DS2 stated: "*There was a misalignment in planning regarding each team's dependen- cies. For example, software engineers sometimes depended on a change in the* *model that was not in our backlog. The roles of each team ended up not being clear, which led to problems in the API used to consume the model. We lacked comprehensive planning that involved both teams more."*

**Data Management.** The findings that emerged from the qualitative analysis related to data management are shown in Figure 4.

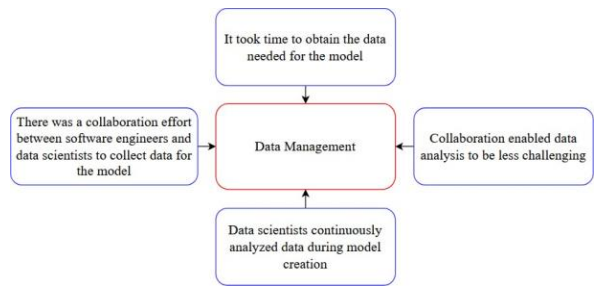

**Fig. 4.** Findings for the Data Management category

**It took time to obtain the data needed for the model**. DS2 described how this situation led to undesired development tasks: "*We also had problems with data availability. It took us some months to get all the valid data neededfor testing. Therefore, we had to initially use mocked data, which later became different from the real data, leading to rework.*"

**There was a collaboration effort between software engineers and data scientists to collect data for the model**. All participants confirmedthe data science team was responsible for analyzing and documenting the data, and no software engineer participated in these activities. Even though software engineers did not directly analyze data, they collaborated with data scientists on other tasks. They were responsible for obtaining the data from customer representatives and making it available to the data science team, as explained by DS1: "*Since we worked with legal processes containing sensitive data, we needed a secure way to obtain them. The development team defined how this would be done together with customer representatives. They created a tool to download the data and make it available on our server.*"

**Data scientists continuously analyzed data during model creation**. Both data scientists agreed that the analysis was not trivial. DS2 explained: "*Pre-processing the data was complex. We received raw data, so cleaning proce- dures were necessary, and we also put a lot of effort into annotating the data.It took a lot of effort to analyze and process the data received so that we could work on the model. This situation also affected what algorithms we could use for the model.*" Data  analysis also uncovered new model input  fields that  needed to be included, as explained by DS1: "*[...] it took us a while to figure out what* *data we needed to request from the customer. We defined some data fields during development, while others were defined during meetings"*.

**Collaboration enabled data analysis to be less challenging**. As raw data was received in files, the software engineers created a tool to help with the analysis. DS1 explained: "*The development team helped us to create a text annotation system and make it available to the domain experts. They [the domain experts] indicated which document parameters were most interesting for extraction and annotated the data we used for model training. They were always by our side to answer questions, which was essential for building the model.*

**Model Management.** Results that emerged from our analysis related to model management are portrayed in Figure 5.

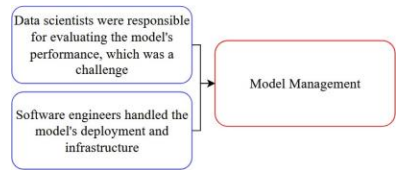

**Fig. 5.** Findings for the Model Management category

**Data scientists were responsible for evaluating the model's performance, which was a challenge**. DS1 and DS2 mentioned that it was not easy to define metrics for the model, such as a target accuracy. DS2 mentioned "*the time taken to obtain valid data hindered the time to create a better performance evaluation framework.*" To solve this, the team agreed to validate the model results together with customer representatives. DS2 explained: "*We presented model studies to the project's stakeholders for them to evaluate if the results were adequate or not.*" DS1 also mentioned that "*a committee of customer representatives was responsible for validating the results produced by the model.*". When asked about implementing incremental learning for the model, participants DS1 and DS2 both said this was a future goal.

**Software engineers handled the model's deployment and infrastructure**. SE1 stated: "*We are responsible for deploying the model consumption API. The deployment of this service [...] is automated through a CI/CD pipeline.*" SE2 explained why the software engineers had this responsibility: "*We already had a pattern for deployment beforehand, and we knew the data scientists did not specialize in DevOps, so we left this structure ready for them.*" Furthermore, the software engineering team developed the web application and the back-end services that consume the model.

**Team Interaction** Finally, Figure 6 presents our findings related to team interaction. Each finding is discussed below.

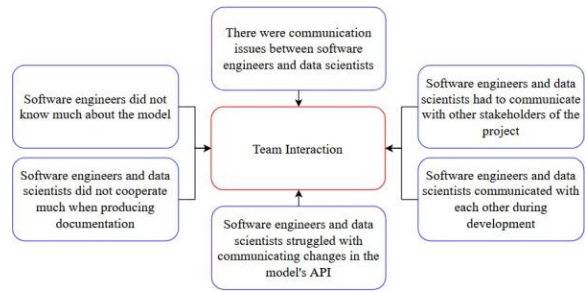

**Fig. 6.** Findings for the Team Interaction category

**There were communication issues between software engineers and data scientists**, which caused problems in the ML-enabled system. SE1 ex- plained the discovery of errors in the data received by the model: "*I did not have the necessary knowledge to analyze if the data was correct and what fields were required or optional. Problems only appeared when we started testing. [...] If the teams had not been so distant, we could have anticipated these problems.*" SE2, on the other hand, exemplified communication issues by explaining how the team should have discussed how to store the model artifacts: "*We provided Git repositories for this storage, but the teams did not discuss how the data scientists would store the artifacts. This eventually caused issues because the model had a lot of artifacts, such as the training scripts, which were not separated from the API code. For this reason, large files were loaded unnecessarily every time a new model release was generated.*"

**Software engineers did not know much about the model**. Since data scientists and software engineers had different responsibilities, they became un- aware of each other's activities. SE1 explained his view of this situation: "*We were very separated, and I did not like that. We did not know much about the model. It was like a 'black box' [...]. Even with a well-defined API, things that were obvious to the data science team were unclear to us. [...] We only developed the services that consumed it, so we did not know what was being done.*" This situation proved to be a problem when defining the model's input, as SE1 high- lighted: "*When we met with customer representatives and data scientists to map the data required by the model, I was unsure if the data we requested was correct since I did not know what the data scientists expected for the model input.*"

**Software engineers and data scientists did not cooperate much when producing documentation**, as each team was responsible for document- ing different parts of the system. DS1 explained: "*The software engineers docu- mented the input data, while we documented the output data.*" However, changes in the system harmed this process, as explained by SE2: "*Our biggest challenge was regarding the changes. The system's initial state was well-documented, but then changes started happening. These changes were not documented properly, which harmed the alignment between the teams. We did not correctly update the* *documentation throughout the project, and we also did not communicate these changes efficiently. We discovered them as system components stopped working.*" Likewise, SE1 was not fully satisfied with the system's documentation: "*It is not documented well enough. We currently have the model's output and input data documented. But, for example, in the middle of this integration, there is a mapper that converts data to the format expected by the model. We could have documented this conversion better.*" SE1, who joined the team in the middle of the project, explained that initially he did not know about documentation that could have helped him during his onboarding: "*I became aware of the model's objectives and the system architecture during the project. I would ask the data science team questions when I had doubts. There was no formal passage of knowledge but instead explanations on demand.*"

**Software engineers and data scientists had to communicate with other stakeholders of the project**. SE2 gave an example: "*We had several discussions with customer representatives to understand their product vision and define what was possible. From there, the UX designers started to prototype ideas that we later used to model the system database*". Data scientists also interacted with customer representatives to define requirements and explain the business rules behind the model's behavior.

**Software engineers and data scientists communicated with each other during development**, and we noticed how they had a good relationship inside the team. DS1 emphasized this by stating: "*We do not have any problems in terms of communication between the teams, as the software engineers are very attentive and available to us. When there is a change, like  new data that needs to be included in the API, or when there is an issue, we communicate directly through messages.*"

**Software engineers and data scientists struggled with communicat- ing changes in the model's API.** SE2 stated: "*Problems in the ML-enabled system were caused by changes in the communication interface established for the API.*" DS2 expressed dissatisfaction with errors in the model's input: "*Problems with input data formats when calling the model's API should not have been our responsibility, as this data had to be in the expected form before communication happened. [...] we had to build workarounds to correct input data formats, which made the system's integration with the model take time and generate rework.*"

## 5   Discussion

### 5.1   How do software engineers and data scientists share responsibilities when developing an ML-enabled system?

Data scientists and software engineers had specific responsibilities in the project. Data scientists focused on analyzing data and developing the model together with its consumption API. Software engineers, on the other hand, were responsible for the model's infrastructure and the back-end services that access it.

Both teams shared responsibilities with other project members and stakeholders. For example, they participated in meetings with customer representa-tives to define the ML-enabled system's goals and functionalities. Data scientists worked closely with domain experts to understand data fields and discover new ones subsequently included in the model's input. Software engineers discussed interface layouts with UX designers before implementing them on the system's web application. These findings are in line with Zhang *et al.*'s work [14], which indicates that software engineers and data scientists are present in different stages of the project, from developing the system to communicating with stakeholders.

Participants illustrated multiple interaction points between the software engineering and data science teams. They had several meetings to define model inputs and outputs and to enable model integration with the rest of the system. The same happened during data collection, when software engineers helped data scientists obtain the data for model training. Both teams also interacted during data annotation, as the software engineers created a system to help with this process. The developed tool allowed domain experts to select text areas inside the files and associate them with a data field, structuring the data for the model.

The interviews revealed that DS2 was not pleased with all the responsibilities his team received. They had to map the business flow behind processing legal disputes, elicit requirements for the model, and present ideas to stakeholders. They also had to make several decisions regarding model features; not all could be validated with customer representatives. DS2 also explained the data science team's participation in developing the model consumption API. Even though they were a research team, they developed all the API's code, a skill they did not have much experience with. For this reason, software engineers helped them during the process. Software engineers also handled the model's infrastructure and deployment pipeline since this was another skill the data scientists did not possess. Data scientists struggling with ML infrastructure was a concern mentioned in Nahar *et al.*'s work [16].

Team members performing activities outside their field of expertise highlights an opportunity to improve planning, which was another topic mentioned during the interviews. Participants revealed that features developed by software engineers could not be deployed because data scientists had to prioritize other functionalities. Although both teams tried to work as independently as possible, having such dependencies effectively mapped and planned could have enhanced the team's delivery speed and avoided problems during the model's integration.

### 5.2 How do software engineers and data scientists collaborate when developing an ML-enabled system?

Participants viewed communication between the software engineering and data science teams positively. Both teams had a good relationship and were always helpful when a member had doubts. They had a group chat where they could interact at any given time.

However, their communication could have been more efficient. Both teams worked almost independently. This reduced the frequency of interactions between them, which led to software engineers not having much knowledge about the model. Even though the model had a well-defined API, which was discussed by both teams, SE1 and SE2 used the term "black box" to describe the ML component. This lack of knowledge became evident during meetings with customer representatives, as there was a mismatch between the participants' understanding of the data. For example, one of the software engineers could not evaluate if the requested data was sufficient for the model, nor if they were in the expected format. This situation caused errors in the ML-enabled system that were only discovered during testing, resulting in avoidable rework.

Constant changes in requirements, also observed in Wan *et al.*'s work [23], worsened the ineffective communication between data scientists and software engineers. As new requirements appeared, the model and the system had to be updated. The data science team had to implement new data fields for the model's input and business rules for the model's output. At the same time, software engineers needed to change the system to capture such data fields, either by user input or through accessing an external API. These changes provoked errors in the system because they were not communicated properly among the teams. For this reason, data scientists had to develop adaptations in the model consumption API to accommodate different input data formats.

The team made an effort to document product definitions and the ML-enabled system architecture. Meeting decisions were registered in minutes, and data scientists and software engineers were responsible for documenting different system components. Software engineers documented the model's input data and the back-end services that consume the model. Data scientists documented the business rules behind the model's behavior and its output responses.

This separation of responsibilities made maintaining the documentation harder. New features were constantly being developed, and the team struggled with keeping documents up to date. The aforementioned inefficient communication of changes in the system was  another obstacle when updating  documentation. For example, problems with changes in the model's input were fixed by creating mappers that corrected the format of input data fields, and one participant mentioned that these mappers could be better documented.

Communication between the data science and software engineering teams was essential for one participant who joined the team after product development had started. SE1's understanding of the ML-enabled system's objectives and architecture was acquired through conversations and questions to the team, as no formal documentation was presented to him. The data fields used by the model are very specific to its domain, which makes understanding them difficult for someone unfamiliar with all the business rules of legal procedures.

## 5.3   Comparison with Literature

Our findings are consistent with results from previous studies regarding the collaboration between data scientists and software engineers developing ML-enabled systems. Many collaboration challenges discussed by Nahar *et al.* [16] were reported in our interviews, such as data scientists working isolated from software engineers, insufficient system documentation, and problems with responsibility sharing. The authors identified three collaboration points: identifying and decomposing requirements, negotiating training data quality and quantity, and integrating data science and software engineering work. All of these points were also present in our case study project. Our findings even reported a new collaboration point, where software engineers developed a system used by domain experts to help in data annotation for model training.

We could also identify several challenges illustrated by Mailach and Norbert [15]. For instance, it was clear that the software engineers did not know enough about the model, describing it as a black box. In addition, we noticed disconnections between the development team and some project stakeholders, especially when defining requirements for the model. In our research, however, the participants did not explicitly mention production delays due to these adversities.

### 5.4 Implications for Practitioners

Based on our findings, we present recommendations for practitioners to improve collaboration between software engineers and data scientists. We seek to aid teams developing ML-enabled systems to avoid the abovementioned pitfalls.

One of the key challenges that software engineers and data scientists face when interacting and collaborating on ML-enabled systems is the lack of clear requirements specifications. Without well-defined requirements, it can be difficult for these actors to understand each other's needs and expectations, leading to miscommunication and inefficiencies in the development process. This highlights the importance of establishing and maintaining clear requirements specifications that can serve as a shared understanding between software engineers and data scientists, enabling them to work together more efficiently.

Fostering a collaborative culture from the start of the project is fundamental. We believe this can be achieved by establishing a comprehensive planning of the system that involves all actors and stakeholders. While planning, the responsibilities of each actor must be clear to everyone on the team. Moreover, actors should be comfortable with the tasks they will perform or at least be willing to learn how to execute them. If there are any dependencies between actors that require their cooperation, these should be mapped in advance to prevent any delays during development.

Despite their background and cultural differences, software engineers and data scientists should avoid working isolated from one another. Even though some tasks can be executed independently, they need to communicate frequently. Teams should also encourage knowledge exchange between them, which can be done by pairing a member from each role to work on a task together. Another possibility is to have members of a role present their work to the rest of the team so that other actors can become familiar with their activities.

ML-enabled system architecture and definitions documentation can also enhance the interaction between these two actors. These documents should provide a concise and unambiguous description of what the ML-enabled system and each of its components should do. This facilitates the discussion between team members, who can use this documentation as a reference, preventing misconceptions. As illustrated by our results, such documentation can also be extremely useful while onboarding new team members.

### 5.5    Threats to Validity

This section discusses threats to validity, focusing on four types of threats: construct validity, internal validity, external validity, and reliability [4].

**Construct validity** refers to whether the applied research methodology is suited to answer our research questions. To mitigate threats, two of the authors had access to project documents, such as use case diagrams and system architecture documents. This documentation was used to cross-check the participants' statements. In addition, all authors revised the transcriptions, codes, and themes generated during the analysis. At the same time, the coding process in RTA is inherently subjective [6], where researchers use their own experiences while interpreting the data.

**Internal validity** is the extent to which our study presents truthful results for our population. To mitigate threats, we formulated the interview questions based on the findings of Villamizar *et al.* [7], which were acquired through a literature review [11] and reports of industrial experiences with ML systems [24]. We also explained the questions in detail when the participants expressed doubts to leave as little room for misunderstandings as possible. We recognize the number of participants, which was limited because of the team's size, may affect the credibility of our results. To mitigate this, we interviewed the team's most experienced software engineers and data scientists.

**External validity** concerns how our findings can be generalized. We understand that our case study only discussed challenges from a single team working in a specific ML-enabled system. It is possible to have scenarios where, for example, the same team is responsible for all tasks carried out by software engineers and data scientists. In other cases, a project manager might define responsibilities more formally, which can alter the team's collaboration procedures. However, given that some of our results are also present in the current literature, we believe that our case study provides additional insights that may be considered when analyzing the interaction between these two actors.

**Reliability** assesses to what extent the study is dependent on the specific researchers. To improve reliability, besides the peer-reviewed qualitative procedures, we uploaded the transcription of each interview to an online repository[3], enabling auditing our analyses and facilitating the replication of our study.

## 6    Concluding Remarks

This paper investigated the interaction between data scientists and software engineers through a case study with a team developing an industry ML-enabled system. We interviewed two experienced members of each role about their activities and collaboration practices. We used RTA to inspect the transcriptions and extract relevant data to answer our research questions. The results gave us an overview of how the team organized their tasks inside the project and the challenges data scientists and software engineers faced. These include actors being unaware of each other's activities, frequent requirement changes, unsynchronized planning, and outdated documentation of the ML-enabled system. Our study provides concrete examples of these challenges based on a case study of a real ML-enabled system development context. These challenges were also mentioned in related work employing different empirical strategies.

---

[3] https://doi.org/10.5281/zenodo.10035304

Understanding how the collaboration between software engineers and data scientists unfolds inside teams with different compositions and companies with other organizational structures can enhance our findings and verify the occurrence of the challenges we reported. Therefore, we invite the community to conduct additional case studies in a variety of contexts to increase external validity. Furthermore, future work could also consider expanding our study focus to collaboration with other roles, such as business stakeholders and domain experts. These actors were continuously cited during the interviews, given their importance in defining requirements and explaining the data.